\documentclass[12pt]{article}
 \usepackage{xcolor}
\usepackage{amsfonts,amsmath}
\usepackage[mathscr]{eucal}
\usepackage{amssymb}
\usepackage{cite}
\usepackage{bm}
\usepackage{bbold}
\usepackage{amsthm}
\usepackage{graphicx,subfigure}
\usepackage{epsf}
\theoremstyle{plain}

\newcommand{\be}{\begin{eqnarray}}

\newcommand{\ee}{\end{eqnarray}}
\newcommand{\nn}{\nonumber \\}
\newcommand{\lb}{\label}
\newcommand{\p}[1]{(\ref{#1})}

\newcommand{\ga}{\lower.7ex\hbox{$
\;\stackrel{\textstyle>}{\sim}\;$}}

\newcommand{\la}{\lower.7ex\hbox{$
\;\stackrel{\textstyle<}{\sim}\;$}}

\newcommand{\dq}{{\dot q}}
\newcommand{\dQ}{{\dot Q}}
\newcommand{\eq}{\eqref}

\begin{document}

\begin{titlepage}

\vspace*{0.2cm}

\renewcommand{\thefootnote}{\star}
\begin{center}

{\LARGE\bf  Dynamical systems with benign ghosts}\\

\vspace{0.5cm}

\vspace{1.5cm}
\renewcommand{\thefootnote}{$\star$}

\quad {\large\bf Thibault~Damour} 
 \vspace{0.5cm}

{\it Institut des Hautes Etudes Scientifiques, 91440 Bures-sur-Yvette, France}\\

\vspace{0.1cm}

{\tt damour@ihes.fr}\\

\vspace{2mm}

\centerline{and}

\vspace{2mm}

\quad {\large\bf Andrei~Smilga} 
 \vspace{0.5cm}

{\it SUBATECH, Universit\'e de Nantes,}\\
{\it 4 rue Alfred Kastler, BP 20722, Nantes 44307, France;}\\
\vspace{0.1cm}

{\tt smilga@subatech.in2p3.fr}\\

\end{center}
\vspace{0.2cm} \vskip 0.6truecm \nopagebreak

\begin{abstract}
\noindent

We consider finite and infinite-dimensional {\it ghost-ridden}  dynamical  systems whose Hamiltonians involve non positive definite kinetic terms. 
We point out the existence of three classes of such systems where the ghosts are {\it benign}, i.e. systems whose 
evolution is  unlimited in time:  
 {\it (i)} systems obtained from the variation of bounded-motion systems; {\it (ii)}  systems describing 
 motions over certain Lorentzian manifolds and {\it (iii)} higher-derivative models related to certain
 modified Korteweg--de Vries equations.
\end{abstract}

\vspace{1cm}
\bigskip

\newpage

\end{titlepage}

\setcounter{footnote}{0}

\setcounter{equation}0
\section{Introduction}

A ghost-ridden dynamical {\it quantum} system is defined as a system whose spectrum is not bounded neither from below, nor from above.
 This is in particular the case for the quantum versions of the Ostrogradsky 
Hamiltonians \cite{Ostro} describing the dynamics of higher-derivative Lagrangians (for a review,
see, e.g., \cite{Smilga:2017arl}).
Many ghost-ridden systems  are sick: their evolution operator is not unitary. Such systems involve classical trajectories that run into a singularity after
a finite time of evolution (a {\it blow-up}).

There are, however,  ghost-ridden systems with {\it benign} ghosts, in the sense
that they exhibit no classical blow-up and have a unitary quantum evolution operator. 
One of the simplest examples of such a benign-ghost system is the {\it Pais-Uhlenbeck oscillator} \cite{PU} with the higher-derivative Lagrangian
 \be
      \lb{LPU}
      L \ =\ \frac 12 \left[ \ddot{x}^2 - (\omega_1^2 + \omega_2^2) \dot{x}^2 + \omega_1^2 \omega_2^2 x^2 \right]. 
        \ee
The quantum version of the corresponding Ostrogradsky Hamiltonian  does not have a bottom. Indeed, there exists a canonical transformation \cite{Man-Dav} that brings the Hamiltonian into the form\footnote{We assumed  here that $\omega_1 \neq \omega_2$. In the case of equal frequencies, the situation is somewhat more complicated because the canonical transformation mentioned above is singular and the Hamiltonian is not reduced to the form \p{H-raznost}. Still, the Hamiltonian is well defined. It has a continuous spectrum \cite{PU,Bolonek,PU-Sigma}.} 
\be
  \lb{H-raznost}
  H = \ \frac {\hat{P}_1^2 + \omega_1^2 X_1^2}2  - \frac{\hat{P}_2^2 + \omega_2^2 X_2^2}2 \, .    \ee
    This Hamiltonian has a pure point spectrum with neither bottom, nor ceiling:
    \be
    \lb{spec-PU}
    E_{nm} \ =\ \left(n + \frac 12 \right) \omega_1 - \left(m + \frac 12 \right) \omega_2 \,.
     \ee
     If the ratio $\omega_1/\omega_2$ is irrational, the spectrum is everywhere dense. The evolution operator is unitary.
     
     The ghosts generally strike back, however, when one leaves the realm of quadratic Hamiltonians
    to include interactions.  Consider for example the Lagrangian
 \be
      \lb{LPU-nonlin}
      L \ =\ \frac 12 \left[ \ddot{x}^2 - 2\omega^2  \dot{x}^2 + \omega^4  x^2 \right] - \frac 14 \alpha
      x^4 \, , 
        \ee
whose classical equation of motion reads 
  \be
  \lb{eqmot-PUnonlin}
   \left( \frac {d^2}{dt^2} + \omega^2 \right)^2  x - \alpha x^3 \ =\ 0 \, .
    \ee 
The classical trajectories depend on four initial conditions. There is an obvious stationary point
 \be
 \lb{stat-point}
 x(0) \ = \ \dot{x}(0) \ = \ \ddot{x}(0) \ = \ x^{(3)}(0)  \ =\ 0\, .
  \ee

  The behaviour of the system in the vicinity of this point depends on the sign of $\alpha$. If $\alpha <0$, the Ostrogradsky Hamiltonian  acquires an extra negative contribution to the energy and all the trajectories other than $x(t) = 0$  run away to infinity in a finite time. 
  The situation is somewhat better  for positive $\alpha$. The stationary point \p{stat-point} lies at the center
  of an ``island of stability'' --- the trajectories with initial conditions sufficiently close to  
  \p{stat-point} do not go astray, but 
  exhibit an oscillatory behavour. However, this island has a shore. When the deviations of initial conditions from \p{stat-point} are large enough, the trajectory blows up \cite{malicious}. Numerical studies show that most systems obtained by a nonlinear deformation of \p{LPU} exhibit  a similar behaviour.
  For instance, a similar island of stability (surrounded by an infinite ocean of blow-up behavior)
  was  observed in \cite{Carroll} for a (cosmology-inspired) model obtained by adding
  to the Pais-Uhlenbeck Hamiltonian \p{H-raznost} (with $\omega_1=0$ and $\omega_2 \neq 0$) 
  an interaction term $\lambda X_1^2 X_2^2$ (with $\lambda >0$). 
  
When considering  the quantum version of ghost-ridden Hamiltonians involving non-trivial interactions
it is difficult to reach definite answers applying to general situations. A case by case study
is required. For instance, it has been argued that  the quantum problem for the system \p{LPU-nonlin} is also malignant \cite{Smilga:2017arl}. On the other hand, there are Hamiltonians
which entail blowing up classical motions but which lead to a well-defined, unitary quantum evolution. 
A well-known example is the Hamiltonian
 describing the 3-dimensional motion of a particle in an attractive $\frac1{r^2}$ potential:
    \be
       \lb{H-center}
       H \ =\ \frac {\vec{p}^2}{2m} - \frac {\kappa} {r^2} \, .
        \ee
 Classically, for certain initial conditions, the particle falls to the center
 in a finite time. The quantum dynamics of this system depends  on the value of ${\kappa}$. If $m{\kappa} < 1/8$, the ground state
 exists and unitarity is preserved.  If $m{\kappa} > 1/8$, the spectrum is not bounded from below and, what is worse, the quantum problem cannot
 be well posed until the singularity at the origin is smoothed out \cite{Popov}. 
 One can say that for $m{\kappa} < 1/8$ quantum fluctuations cope successfully with the attractive force of the potential and prevent the system from collapsing.
 
The latter example suggests that quantum fluctuations can only make a 
ghost-ridden system better, not worse. 
We therefore {\it conjecture} that, if the classical dynamics of the system is benign, 
its quantum dynamics will also be  benign, irrespectively of whether the spectrum has, or does not have, a bottom.
 
Some particular examples of benign ghost-ridden (nonlinearly interacting) 
Hamiltonians have been presented in previous works \cite{malicious,Robert,Pavsic,Kovner,duhi-v-pole,Deffayet:2021nnt}. 
The aim of the present paper is to delineate new classes of
benign Hamiltonians, some of which have a large generality (in the sense that they
contain several arbitrary functions). More precisely, we shall present classes of nontrivially interacting
ghost-ridden systems such that all the {\it classical} motions (and not only the motions restricted to a
 limited region of phase space) admit an infinite-time evolution.\footnote{Note that we are not requiring
 that the motions indefinitely stay within a compact region of phase space. We do not exclude power-law,
 or exponential, runaway behaviors. We are simply excluding finite-time blow-up.} 
 In view of the conjecture stated in the last paragraph, we expect that the quantum dynamics
 of our benign ghost-ridden systems will be well posed. We leave studies of this issue to future work.

 \section{Variational dynamics: a large class of benign ghosts} \label{sec2}
  \setcounter{equation}0
 
 A general class of dynamical systems with benign ghosts is obtained by considering the 
 variational equations of motion of a bounded-motion\footnote{Here, ``bounded motion" 
 means that the time evolution of the considered, unperturbed, system stays within some compact
 domain of phase space.} Hamiltonian system. An example of such a system was 
 found in \cite{Robert}  by studying a certain higher-derivative supersymmetric mechanical system,
 though its variational nature was not noticed.\footnote{The supersymmetric aspects of this problem
 are not relevant here, and we shall forget here about fermions.} 
 
 The general setting for defining such systems is the following. One starts from a basic, {\it unperturbed}
 dynamical system (with $n$ degrees of freedom) described, say, by a Lagrangian, $L_0(q^i, \dq^i)$, 
 $i=1,\cdots n$, i.e. by the action
 \be \label{S0}
 S_0[q,\dq] = \int dt\,  L_0(q^i, \dq^i)\,.
 \ee
 Then, one considers the dynamics defined by varying the Lagrangian action \eq{S0}, i.e. by
 making in $L_0(q^i, \dq^i)$ the replacement
 \be \label{variation}
 q^i(t) \to  q^i(t) + \epsilon \, Q^i(t) \,,\nonumber\\
  \dq^i(t) \to  \dq^i(t) + \epsilon \, \dQ^i(t) \,,
 \ee
 and by keeping {\it only} the term linear in $\epsilon$. In other words, one is considering the new action
 (with $2 n$ degrees of freedom)
 \be \label{S1}
  S_1[q,\dq ; Q, \dQ] = \int dt L_1(q^i, \dq^i ;Q^i, \dQ^i)\,,
 \ee
 where
 \be \label{L1}
  L_1(q^i, \dq^i ;Q^i, \dQ^i) = Q^j \frac{\partial L_0(q,\dq)}{\partial q^j} + \dQ^j \frac{\partial L_0(q,\dq)}{\partial \dq^j}\,.
 \ee
 The variational action \eq{S1} leads to the following equations of motion for $q^i$ and $Q^i$ 
 (with $\delta/\delta q$ denoting an Euler-Lagrange variational derivative, 
 $\partial_q - \frac{d}{dt} \partial_{\dq}+ \frac{d^2}{dt^2} \partial_{\ddot q} +\cdots$):
 \be
 &&0=\frac{\delta L_1}{\delta Q^i}=  \frac{\partial L_0(q,\dq)}{\partial q^i}- \frac{d}{dt}\frac{\partial L_0(q,\dq)}{\partial \dq^i} \equiv \frac{\delta L_0}{\delta q^i} \,,\nonumber\\
 &&0=\frac{\delta L_1}{\delta q^i}= Q^j \frac{\partial L_0(q,\dq)}{\partial q^j \partial q^i}
 - \frac{d}{dt} \left[ \dQ^j \frac{\partial L_0(q,\dq)}{\partial \dq^j \partial \dq^i} \right]\,.
 \ee
We see that the first Euler-Lagrange equation 
 $0 = {\delta L_1}/{\delta Q^i}$ coincides with the unperturbed equation
 of motion of $q^i$,   ${\delta L_0}(q,\dq)/{\delta q^i} = 0$.
 On the other hand, the second equation, $0 = {\delta L_1}/{\delta q^i}$
coincides with the {\it variation} of the equation ${\delta L_0}(q,\dq)/{\delta q^i} = 0$: it is obtained by replacing there $q^i(t)$ and their first and second derivatives as in \eq{variation} and keeping only the  term linear in $\epsilon$.

 When considering the first-order Hamiltonian version of the so-defined variational dynamics,
 one encounters a slightly surprising feature. Namely, if we denote the momenta conjugate
 to $q^i$ and $Q^i$ respectively as  $p_i$ and  $P_i$, the Hamiltonian
 $ H_1(q^i, Q^i; p_i, P_i)$ describing the varied dynamics (corresponding to
 $L_1(q^i, \dq^i ;Q^i, \dQ^i)$) is {\it not} obtained by varying the unperturbed
 Hamiltonian $ H_0(q^i; p_i)$ by means of the naively expected variation
 $ q^i \to  q^i + \epsilon \, Q^i ;   p_i \to   p_i + \epsilon \,  P_i$.
 It is obtained by first doing the latter naive variation, and then by swapping the
 momenta according to $  p_i \leftrightarrow P_i$.

  The necessity to swap $ p \leftrightarrow P$ in the naive variation of $ H_0(q^i; p_i)$ is easily seen by varying  the Hamiltonian version of the action \eq{S0}, 
 \be \label{S0H}
  S_0^H[q,p] = \int dt \left[ p_i \dq^i - H_0(q,p) \right]\,.
 \ee
 Using the naively defined variations  $ q^i \to  q^i + \epsilon \, Q^i ;   p_i \to   p_i + \epsilon \,  P_i$
 we get
 \be
  \delta S_0^H[q,Q,p,P] = \int dt \left[ P_i \dq^i + p_i \dQ^i - \left( Q^i \frac{\partial H_0(q, p)}{\partial q^i}+  P_i \frac{\partial H_0(q, p)}{\partial p_i}\right)\right]\,,
 \ee
 in which the Hamiltonian kinetic term $P_i \dq^i + p_i \dQ^i $ shows that the conjugate momentum
 to  $q^i$ is actually $P_i \sim \delta p_i$, while the conjugate momentum to $Q^i$ is the unperturbed $p_i$.
After the swap $ p \leftrightarrow P$, one finally gets the  Hamiltonian for the varied dynamics expressed in terms of
the canonical pairs $(q,p); (Q,P)$:
  \be \label{H1}
  H_1(q, Q; p, P)= 
  Q^i \frac{\partial H_0(q, P)}{\partial q^i}+ p_i \frac{\partial H_0(q, P)}{\partial P_i}\,. 
 \ee

 From our present perspective (namely, studying ghost-ridden dynamics), note that all the
 varied Hamiltonians $ H_1(q,Q; p, P)$ are necessarily unbounded below (and above)
 because they have a linear dependence on the phase-space variables $Q^i$ and $p_i$,
 see Eq. \eq{H1}.  This makes the quantum spectrum also unbounded.
 Let us, however, see why this ghost feature    leads generically to a benign evolution.
 
 The simplest type of varied dynamics is obtained by varying a simple Hamiltonian 
 (with one degree of freedom) of the form
 \be \label{H0V}
 H_0(x,p) = \frac12 p^2 + V(x)\,.
 \ee
 In that case, the variation of the first term yields the symmetric structure $pP$, invariant under  the swap  $ p \leftrightarrow P$, which is thus  ineffective. One then gets
 (with the notation $D \equiv Q = \delta x$  used in \cite{Robert}) a varied-dynamics Hamiltonian involving
  two pairs of canonically conjugated  variables, $(x, p)$ and $(D, P)$:
 \be \label{H-Robert}
 H_1(x,D;p,P) = p P + D V'(x)\,.
 \ee
The classical equations of motion are 
   \be
   \lb{eqmot-Robert}
   \ddot{x} + V'(x) \ =\ 0\,, \ \ \ \ \ \ \ \ \ \ \ \ddot{D} + V''(x) D = 0 \, .
    \ee
 They admit two constants of motion: the varied Hamiltonian $ H_1(x,D;p,P)$,
 and the conserved energy $E_0=\frac12 \dot x^2 + V(x)$ of the 
unperturbed motion of $x$.
 Using the Hamilton equation $\dot x = \frac{\partial H_1}{\partial p}= P$, the unperturbed
 energy $E_0(x , \dot x)$ yields the second integral of motion $N(x,P)$ with
   \be
        \lb{N}
        N (x,P)\ =\ \frac {P^2}2 + V(x) \, .
          \ee
 The model of Ref. \cite{Robert} was of this type, with the simplest nontrivial  potential 
  \be
      \lb{V-Robert}
      V(x) \ =\  \frac {\omega^2 x^2}2 + \frac {\lambda x^4}4 \, , \ \ \ \ \ \ \lambda > 0
      \, .
         \ee
If one introduces the variables 
  \be
  \lb{X12-Robert}
  X_{1,2} \ =\  \sqrt{\frac \omega 2} x \pm  \frac 1 {\sqrt{2\omega}}  D \, , \ \ \ \ \ \ 
   P_{1,2} \ =\  \frac 1 {\sqrt{2\omega}} p \pm   \sqrt{ \frac \omega 2}  P \,,
    \ee
    the Hamiltonian \p{H-Robert}  acquires the form \cite{Robert}
      \be
      \lb{H-raznost-Robert}
      H_1 \ =\ \frac{P_1^2 + \omega^2 X_1^2}2 - 
      \frac{P_2^2 + \omega^2 X_2^2}2 + \frac {\lambda}{4\omega} (X_1 - X_2) (X_1 + X_2)^3 \, .
       \ee
       In other words, this is the
Hamiltonian \p{H-raznost} with degenerate frequencies, 
       with an additional quartic interaction of a special form.
       
       A nice distinguishing feature of the system \p{H-Robert} is its integrability. Indeed, we have
       two degrees of freedom, and two commuting integrals of motion: $H_1$ and $N$. (The
       vanishing of the   Poisson bracket $\{H_1, N\}$ is easily checked.)
       This allows one to find, for  the simple potential \eq{V-Robert}, the solution analytically.
       
First,  the unperturbed dynamics for $x$ describes oscillations in the quartic potential \p{V-Robert}.
The solutions are elliptic functions whose parameters depend on the integral of motion $N$ :
      \be
      \lb{cn}
      x(t) \ =\ x_0 \, {\rm cn} [\Omega(t-t_0), k] \,,
       \ee
where  ${\rm cn}$ is a Jacobi elliptic function (with elliptic modulus $k = \sqrt{m} $), and
       \be \label{cnparams}
       \alpha = \frac {\omega^4}{\lambda N}, \ \ \ 
       \Omega = [\lambda N(4+\alpha)]^{1/4},  \nn
        k^2 \equiv m  = \frac 12 \left[ 1 - \sqrt{\frac \alpha{4 + \alpha}} \right], \ \
        x_0 = \left( \frac N\lambda \right)^{1/4} 
       \sqrt{\sqrt{4+\alpha} - \sqrt{\alpha}}   \, .    
  \ee
  Bearing in mind \p{cn},  the equation for $D$ is a Hill equation  describing an oscillator
 with periodically varying frequency. The solutions of generic Hill equations
 are obtained by Floquet theory and, depending on parameters, can be either
 bounded for all times, or exponentially growing. However, we are here in a special
 case. We know that  the equation for $D$ describes the infinitesimal variations
 of the general solution of the quartic oscillator describing the $x$ dynamics. 
 Therefore we can simply obtain the general solution for $D$ by varying the integration
 constants entering the general solution \eq{cn}. It is enough to get two independent solutions.
 A first one can be obtained by varying the integration constant $t_0$ in Eq. \eq{cn}. This yields
 \be
 \lb{D1}
 D_1(t) \ =\ \frac{\partial [x(t)]}{\partial t_0}=  - \dot{x}(t) = + \Omega x_0 \, {\rm sn}[\Omega t, k] {\rm dn}[\Omega t, k] \,,
   \ee
 where we have set  $t_0 = 0$ after variation. This solution is periodic in $t$.
 
\begin{figure}[h]
   \begin{center}
 \includegraphics[width=.65\textwidth]{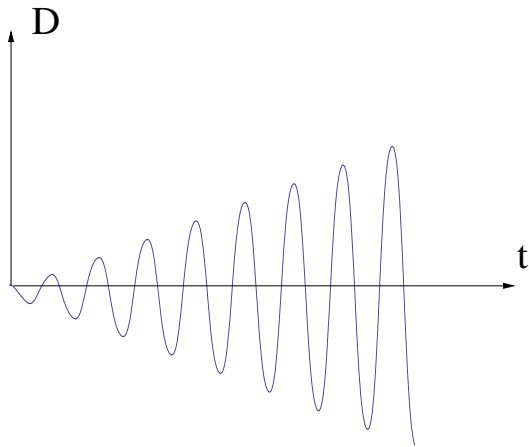}
    \end{center}
\caption{A typical behaviour of $D(t)$, as follows from solving Eq. \p{eqmot-Robert}.} 
\label{Drost}
\end{figure}

 A second solution is obtained by varying the value of $N$ in Eq. \eq{cn}.
 Varying $N$ implies that both the amplitude $x_0(N)$, and the frequency $\Omega(N)$,
 of $x(t)$ must be varied. The variation of  $\Omega(N)$
 in $x(t) \ =\ x_0(N) \, {\rm cn} [\Omega(N) t, k] $ (setting again $t_0=0$,
 and noticing that $k$ does not depend on $N$)  brings about a prefactor $t$ multiplying the derivative of ${\rm cn}(x)$
 and generates a  $D_2(t)$ of the form
 \be
 D_2(t) \ = \ t  \frac{d}{ dt}  {\rm cn} [\Omega(N) t, k] + \beta \, {\rm cn} [\Omega(N) t, k]\ = \nn
- t \,\Omega(N)\, {\rm sn}[\Omega(N) t, k] {\rm dn}[\Omega(N) t, k] +   \beta\,  {\rm cn} [\Omega(N) t, k], 
 \ee
 where $\beta$ is some constant. The contribution $- t \Omega \, {\rm sn}[\Omega t, k] {\rm dn}[\Omega t, k]$
 is the product of $t$ by a periodic function of $t$. Therefore, while the first independent solution
 $D_1(t)$ is periodic, the second independent solution  $D_2(t)$ will exhibit an oscillatory
 behavior with an amplitude rising linearly in time. This is illustrated in Fig. \ref{Drost}.
What is important for our present purpose is the benign nature of the
general solution, $D(t)= c_1 D_1(t) + c_2 D_2(t)$.
 Indeed, we consider as malignant ghost only the cases leading
to blow-up in a finite time.  The linear growth in time exhibited by the generic solution of
the $D(t)$ equation is quite benign.
     
We refer to Ref. \cite{Robert} for a study of the quantum version of the Hamiltonian \eq{H-Robert}
(due to integrability, the  eigenvalues and eigenfunctions of the quantum Hamiltonian can be found explicitly in this case).

We have used here the specific quartic potential \eq{V-Robert} to be able to exhibit explicit solutions
for the one-degree-of-freedom variational dynamics \eq{H-Robert}. However, the conclusions we reached
about the benign dynamics of the $(x,D;p,P)$ system hold for a general class of potentials.
Indeed, if we take for $V(x)$ a smooth confining potential growing as $ x \to \pm \infty$, 
the solutions for $x(t)$ will  represent a nonlinear oscillation  of a certain type 
--- periodic functions of time with a  frequency $\Omega$ 
depending on the integral of motion 
\be
E_0(x , \dot x)\equiv\frac12 \dot x^2 + V(x)= \frac {P^2}2 + V(x) \equiv N\,.
\ee
 In other words,
\be
x(t) = X\left[\Omega(N) (t-t_0), N \right]\,,
\ee
where $X(\theta ,N)$ is a periodic function of $\theta$, with period $2 \pi$. It  can be expanded into a Fourier series,
\be
X(\theta,N)= {\rm Re} \left\{\sum_{k \in  \mathbb{Z}} A_k(N) e^{{\rm  i } k \theta} \right\}\,.
\ee
The existence of the two commuting integrals of motion $H_1$ and $N$ allows one to solve
the motion of $x(t)$ by a quadrature, giving some type of  hyperelliptic function.

 The general solution for  $D(t)$ can then be 
simply obtained by varying the general solution for $x(t)$ with respect to the two integration
constants it contains, namely $t_0$ and $N$. Again, it is clear that the first independent solution
\be
D_1(t) = \partial x(t)/\partial t_0= - \dot x(t)=\Omega(N){\rm Im} \left\{\sum_k k A_k(N) e^{{\rm  i } k \theta(t)} \right\}\,,
\ee
will be periodic in time (with the same
period as $x(t)$), while the second solution
\be
D_2(t) = \frac{\partial}{\partial N} X[\Omega(N) (t-t_0), N]= {\rm Re}  \frac{\partial}{\partial N} \left\{\sum_k A_k(N) e^{{\rm  i } k \Omega(N) (t-t_0)} \right\}\,,
\ee
will be the sum of a periodic function, and of the function
\be
\frac{d \ln \Omega(N)}{dN}(t-t_0) \dot x(t)\,.
\ee
 The latter function is a product of $t-t_0$ and of a periodic function of $t$. As in the case
 illustrated in Fig. \ref{Drost}, we have again an oscillatory
 behavior with an amplitude rising linearly in time. A benign ghost again.

Let us consider the generalization of these results   to $n$ degrees of freedom, i.e. to the system with the varied
Lagrangian \eq{L1} and the corresponding Hamiltonian \eq{H1}. 
In that case, the results will strongly depend
on the integrable or non-integrable character of the unperturbed dynamics described by 
the action \eq{S0}, or equivalently, \eq{S0H}.

Let us first assume that the unperturbed dynamics \eq{S0H} is {\it integrable}, and that we are considering
bound motions. In that case, the general solution of the dynamics \eq{S0H} corresponds to a 
{\it quasi-periodic} motion where the phase-space coordinates $q^i, p_i$ admit representations of
the form
\be \label{quasiperiodic}
q^i(t)  &=& q^i[ {\bf I}, {\boldsymbol \theta}(t)]=\sum_{\bf k} a^i_{k_1 \cdots k_n}(I_1,\cdots, I_n) e^{{\rm  i } (k_1 \theta_1(t)+ \cdots + k_n \theta_n(t) )}\,, \nonumber\\
p_i(t)  &=& p_i[ {\bf I}, {\boldsymbol \theta}(t)]=\sum_{\bf k} b^i_{k_1 \cdots k_n}(I_1,\cdots, I_n) e^{ {\rm  i } (k_1 \theta_1(t)+ \cdots + k_n \theta_n(t) )}\,.
\ee
Here $({\bf I}, {\boldsymbol \theta})=( I_i, \theta_i)$, $i=1,\cdots,n$ are action-angle variables, 
${\bf k}= (k_i), i=1,\cdots,n$ are multiplets of (relative) integers (summed over $\mathbb{Z}^n$), 
and  the time evolution of the angles is of the form
\be
 \theta_i(t)=  \omega_i({\bf I}) \, t + \theta_i^0 \,.
\ee
We assume that the $2n$ integration constants entering this solution are  $I_i$, and $\theta_i^0$. The general solution
for the variation $Q^i= \delta q^i$ can then be obtained as a linear superposition of the $2n$ particular 
varied solutions defined by varying the $2n$  integration constants, i.e.
\be
Q^i(t)=\sum_j \left[C_{I_j} \frac{\partial }{\partial I_j} q^i[ {\bf I}, {\boldsymbol \theta}(t)] + C_{\theta_j}
\frac{\partial }{\partial \theta^0_j} q^i[ {\bf I}, {\boldsymbol \theta}(t)] \right]\,.
\ee
This expression is the sum of some quasi-periodic functions (of the type \eq{quasiperiodic})
and of the functions coming from varying the frequencies $\omega_i(I_j)$, namely
\be \label{tquasiperiodic}
t \, \sum_{j, \ell}  C_{I_j} \frac{\partial \omega_{\ell}}{\partial I_j}\frac{ \partial  q^i[ {\bf I}, {\boldsymbol \theta}]}{\partial \theta_{\ell}}\,.
\ee
The latter functions are the products of $t$ and quasi-periodic functions of time.\footnote{Here we assume
a fast enough decay for the coefficients $ a^i_{k_1 \cdots k_n}$ as the $k_i$ tend to infinity to ensure
the  quasi-periodic nature of the right factor in Eq. \eq{tquasiperiodic}.} This is again a benign behavior
of linearly-growing oscillatory form, which is essentially a quasi-periodic version of Fig. \ref{Drost}.

 Let us now consider the generic case where the unperturbed system \eq{S0} is {\it not integrable}, and exhibits a chaotic behavior. [We are still assuming that the (unperturbed)
 evolution stays within a compact domain of phase-space.] In that case, the behavior of the varied
 dynamical system can be much worse than in the integrable case assumed above.
 First, on mid-term time scales, the variations $Q^i= \delta q^i$ 
 may grow exponentially with time (Lyapunov instability). 
 On longer time scales, such a Lyapunov exponential instability might (via an Arnold-type diffusion)
 evolve into a more chaotic behavior. Anyway, $Q^i(t)$ satisfies an homogeneous {\it linear} ordinary 
 differential equation (ODE) with time-dependent coefficients, say (in the simple, equal-mass, potential case)
 \be
 \ddot Q^i + \frac{\partial^2 V(q)}{\partial q^i \partial q^j} Q^j=0\,,
 \ee
 in which $q^i$ must be replaced by a solution of  $ \ddot q^i + \frac{\partial V(q)}{\partial q^i }=0$.
 General theorems about linear ODE's then guarantee that $Q^i(t)$ can only have singularities at a finite time,
 if the coefficients of the ODE become singular. As we assume here that the unperturbed solution
 $q^i(t)$ is regular for all times, we are guaranteed that the behavior of $Q^i(t)$ will be
 benign in our general sense (i.e. no finite-time blow-up).

 Let us finally mention that our variational approach can be  
straightforwardly generalized to field theory systems (i.e. to an infinite number of dynamical variables). 
For instance, if we start with the
unperturbed Lagrangian (using here 
the signature $(x_\mu)^2 \equiv t^2 - {\bf x}^2$),
\be
{\cal L}_0 \ = \ \frac 12 (\partial_\mu \phi)^2 - V(\phi)\,,
\ee
we are led to the following variational Lagrangian for two real fields $\phi(t,{\bf x})$ and $D(t,{\bf x}) \equiv \delta \phi(t,{\bf x})$
  \be
  \lb{L-pole}
{\cal L} \ =\ \partial_\mu \phi \partial_\mu D - D V'(\phi). 
  \ee
If we use the same potential as in \p{H-Robert}, the equations of motion for the system \p{L-pole} are
   \be
\lb{eqmot-pole}
\Box \phi + \omega^2 \phi + \lambda \phi^3 &=& 0 , \nn
\Box D + (\omega^2 + 3\lambda \phi^2) \, D&=& 0 \, .
 \ee
 The particular case of a (1+1)-dimensional spacetime has been considered (and numerically
 investigated) in Ref. \cite{duhi-v-pole}.
 
General mathematical results on the nonlinear Klein-Gordon equation (with potential $ \lambda \phi^p/p$,
with $\lambda >0$ and $ p<2+ \frac{4}{d-2}$ in $d$ spatial dimensions) \cite{GinibreVelo1989} guarantee
the global existence  of solutions of Eq. \eq{eqmot-pole} for $\phi(t, {\bf x})$,
with suitable (finite-energy) Cauchy data at $t=0$. These generic solutions have been shown to 
exhibit asymptotic decay in time. We then expect that these exact results on the nonlinear
Klein-Gordon equation for $\phi$ imply (by varying the Cauchy data)
a benign behavior (with asymptotic decay) for generic solutions of the linear, varied Klein-Gordon
equation satisfied by $D=\delta \phi$. Ref. \cite{duhi-v-pole} numerically investigated the case
of a one-dimensional space ($d=1$) compactified on a circle. The space compactification suppresses the
asymptotic time decay of $\phi$ and seemingly induced a mild (linear in time) growth for 
$D$ \cite{duhi-v-pole}.

  Evidently our variational construction can be set up basically for any system. One can vary the Yang-Mills Lagrangian,
     \be
     \lb{YM}
     {\cal L}^{\rm YM}_0 \ =\ - \frac 12 {\rm Tr} \{ F_{\mu\nu} F^{\mu\nu} \}\,,
      \ee
setting $A_\mu \to A_\mu + \epsilon B_\mu$ and  keeping in \p{YM} the terms linear in $B_\mu$, namely
\be
 {\cal L}^{\rm YM}_1 = B_\mu \frac{\delta {\cal L}^{\rm YM}[A]}{\delta A_\mu} \,.
\ee
At the linearized level (in $A$), ${\cal L}^{\rm YM}_1$ describes two massless spin-1 fields, one of which
is a ghost. At the nonlinear level, the number of degrees of freedom is preserved because of the
invariance of ${\cal L}^{\rm YM}_1$ under two distinct gauge transformations: the usual one
acting on $A$, and a separate one (involving $A$-covariant derivatives) acting on $B$.
This system is again a benign ghost-ridden system.

Mutatis mutandis, one can also consider the ghost-ridden action obtained by varying the Einstein action
[we use here
 the signature $(-+++)$ and $16 \pi G=1$],
\be
 {\cal L}^{\rm E}_0=  \sqrt{-g} g^{\mu \nu} R_{\mu \nu}[g]\,.
\ee
Varying $g_{\mu \nu} \to g_{\mu \nu} + \epsilon h_{\mu \nu}$ and keeping the term linear in $\epsilon$ yields
\be
{\cal L}^{\rm E}_1= - \sqrt{-g} \left(  R^{\mu \nu}[g] - \frac12 g^{\mu\nu} R[g]  \right)  h_{\mu \nu}\,.
\ee
At the linearized level around flat spacetime ($g_{\mu \nu} = \eta_{\mu \nu}+ \epsilon f_{\mu \nu}$), ${\cal L}^{\rm E}_1$ describes two massless spin-2 fields, one of which
is a ghost. At the nonlinear level, the number of degrees of freedom is preserved because of the
invariance of ${\cal L}^{\rm  E}_1$ under two distinct gauge transformations: the usual coordinate
transformations, $x'^{\mu} = f^\mu(x^\nu)$, 
acting both on $g_{\mu\nu}$ and on the tensor $ h_{\mu \nu}$, and a separate one
 (involving $g$-covariant derivatives) acting linearly on $h$, namely $ h_{\mu \nu} \to h_{\mu \nu} + \nabla^g_{\mu} \xi_{\nu}+ \nabla^g_{\nu} \xi_{\mu}$. Note that the Euler-Lagrange
 equations derived from ${\cal L}^{\rm E}_1$ imply that $g_{\mu \nu}$ must satisfy
 the (vacuum) Einstein equations, 
 \be
 {\cal E}^{\mu \nu}[ g_{\alpha\beta}] \equiv \sqrt{-g} \left(  R^{\mu \nu}[g] - \frac12 g^{\mu\nu} R[g]  \right)=0\,,
 \ee
 while $h_{\mu \nu}$ must satisfy the {\it linearized} Einstein equations
 \be
  \frac{\delta  {\cal E}^{\mu \nu}}{\delta g_{\alpha\beta}}\, h_{\alpha\beta} = 0\,.
 \ee
This system is again a benign ghost-ridden system. For instance, the mathematical results on the
global nonlinear stability of the Minkowski spacetime 
\cite{Christodoulou:1993uv,Lindblad:2004ue,Bieri:2009xc} have shown
the global existence  of solutions of Einstein vacuum equations for $g_{\mu \nu}(t, {\bf x})$,
with small Cauchy data at $t=0$. These generic solutions have been shown to 
exhibit asymptotic decay in time. We believe that the good control (in all spacetime directions)
of the geometric properties of these nonlinear, but small perturbations of Minkowski space 
\cite{Christodoulou:1993uv} suffice to prove the global existence of the Green's function
needed to solve the linearized Einstein equation satisfied by $h_{\mu \nu}$ (after a suitable
gauge fixing).

The basic reason why the ghost-ridden systems considered in this section were benign
is that the equations of motion of the variables $(Q,P)$ were linear [though influenced by
the nonlinear   dynamics of the unperturbed variables $(q,p)$].  In the following, we are going
to introduce more interesting ghost-ridden systems where the dynamics of  the ghost degrees of
freedom is nonlinear. It is then more delicate to delineate ghost-ridden systems that stay benign
in spite of such nonlinear interactions.

\section{Geodesics on Lorentzian manifolds} \label{sec3}
 \setcounter{equation}0

A general class of nonlinear ghost-ridden Hamiltonians with benign solutions are the (quadratic) Hamiltonians
describing geodesic motion on  {\it geodesically complete Lorentzian manifolds}. 
Namely, to any given  $D$-dimensional
Lorentzian 
 manifold $M_D$ with metric tensor 
$g_{\mu \nu}(x)$ one can associate the Hamiltonian
\be \label{p2}
H(x^\lambda, p_\lambda) = \frac12 g^{\mu \nu}(x) p_\mu p_\nu \,.
\ee
Because of the $- + \cdots +$ signature that we choose to use here, this Hamiltonian contains $D-1$ positive terms and   a  ghost-like negative term. The Hamiltonian \eq{p2} describes geodesic motions on $M_D$. More precisely,
the Hamilton equations of motion
\be \label{geod}
 \frac{d x^\mu}{d \tau} &=&
 \frac{\partial H}{\partial p_\mu}= g^{\mu \nu}(x)  p_\nu\,, \nn
 \frac{d p_\mu}{d \tau} &=& -  \frac{\partial H}{\partial x^\mu}= -  \frac12 \partial_\mu g^{\nu \lambda}(x) p_\nu p_\lambda\,,
\ee
describe a motion in the $2D$-dimensional phase-space  $(x^\lambda, p_\lambda)$ [the cotangent space
of $M_D$] with respect to an Hamiltonian ``time variable" $\tau$, which is an {\it affine} parameter
along the considered geodesic. For a general curved spacetime, the only conserved quantity of the
dynamics \eq{geod} is the ``energy" (the factor 2 being introduced for convenience),
\be
E = 2 \,H(x,p) \equiv  g^{\mu \nu}(x) p_\mu p_\nu\,.
\ee
A positive value of $E$ describes spacelike geodesics, a negative value describes timelike ones, 
while $E=0$ describes null geodesics. Actually, as the affine parametrization of geodesics is defined
modulo an arbitrary affine transformation $\tau \to a \tau + b$, one can, without
loss of generality only consider the three cases $E=+1$, $E=-1$ and $E=0$. The values $E=\pm1$
mean that $\tau$ is equal to the proper length $\sqrt{\pm ds^2}$ along the geodesic, with
\be
ds^2= g_{\mu\nu}(x) dx^\mu dx^\nu\,.
\ee
 We are interested in the systems for which
 the Hamiltonian evolution \eq{geod} can be continued indefinitely
with respect to the Hamiltonian time variable $\tau$. Then, as was argued in the Introduction,  the ghosts are benign. In the context of the geodesic Hamiltonian \eq{p2},
our condition boils down to saying that the Lorentzian manifold $(M_D, g)$ is
{\it geodesically complete}. We can therefore conclude that the ghost-ridden Hamiltonian \eq{p2}
defines a benign dynamics (for all values of the energy $E$) on any geodesically complete Lorentzian
manifold.

Mathematical investigations have given large classes of geodesically complete Lorentzian
manifolds. Of particular physical significance is the fact, proven in 
Refs. \cite{Choquet-Bruhat:2006nor,Loizelet2009}, that the vacuum Einstein spacetimes
close to Minkowski that were constructed in Refs. \cite{Christodoulou:1993uv,Lindblad:2004ue,Bieri:2009xc}
are geodesically complete 
for all values of the energy. Evidently, this property will not extend if one
considers spacetimes containing black holes. 

As a very particular type of geodesically-complete spacetimes, we can also mention
 the de Sitter, as well as anti-de Sitter (AdS), spacetimes (of any dimension) \cite{Hawking:1973uf}.  
For concreteness, let us consider the AdS spacetime, with  the metric (in global coordinates)
\be \label{ads}
ds^2=- \left(1 +  \frac{r^2}{\ell^2}\right) dt^2 + \frac{dr^2}{1 + \frac{r^2}{\ell^2}} + r^2 d\Omega^2_{D-2}\,.
\ee
The associated geodesic Hamitonian reads
\be
H= -\frac12 \frac{p_t^2}{1 +  \frac{r^2}{\ell^2}}+ \frac12 \left(1 +  \frac{r^2}{\ell^2}\right) p_r^2+ \frac12 \frac{J^2}{r^2}\,.
\ee
Here, $\ell$ is the AdS length scale  giving   the constant negative curvature $K = - 1/\ell^2$
and $J^2$ is the squared angular momentum linked to the motion on the sphere $S_{D-2}$.

The AdS spacetime is a homogeneous, symmetric space, equivalent to the coset $O(2,D-1)/O(1,D-1)$.
Because of its homogeneity we can reduce the study of geodesics to the geodesics starting from
any given point, say the origin $t=0, r=0$ in the global coordinates of \eq{ads}. In addition, as the
isotropy group of the coset $O(2,D-1)/O(1,D-1)$, namely $O(1,D-1)$ is the local Lorentz group,
we can use  
 this group to reduce the study of the three types of geodesics to 
 a particular timelike geodesic (e.g. the geodesic $r=0$),  a particular spacelike one
(e.g. the radial geodesic $t=0, \Omega= {\rm cst}$), and a particular null  geodesic
   (a radial one directed along the light cone). 

In these cases, the solutions to the equations \p{geod} are very simple:\footnote{To simplify them still further, 
we have set $\ell = 1$.} the spacelike geodesic with $E=1$ is
$r(\tau) = \sinh \tau, t=0, \Omega= {\rm cst}$, the timelike geodesic with $E = -1$ is $r=0, t = \tau, \Omega= {\rm cst}$,  while the null geodesic is $r= \tau$, $t = {\rm arctan} \,\tau$,  $\Omega= {\rm cst}$.
 We see that all three types of geodesics have an infinite affine length.

The class of Ricci-flat, near-Minkowski geodesically complete Lorentzian manifolds mentioned
above \cite{Choquet-Bruhat:2006nor,Loizelet2009} might naively suggest that any
Lorentzian manifold with small enough curvature will be geodesically complete
 (as is the case  for Riemannian manifolds). In other words, one might think that the incompleteness
of geodesics must be linked to the presence of  curvature singularities, or at least high-curvature regions.
This is incorrect. The  (toy-Taub-NUT) Misner manifold \cite{Misner} 
(see also section 5.8 in \cite{Hawking:1973uf})
yields a simple example of a smooth Lorentzian
manifold, which is {\it locally flat}, but geodesically incomplete (because of its nontrivial topology). 
The Misner manifold is two-dimensional and has  the topology ${\mathbb R} \times S_1$. In global
coordinates $(x,\phi)$, where $x \in {\mathbb R}$ and where $\phi \in [0,2\pi]$ is an angle describing 
the circle $S_1$, the metric reads \footnote{See \cite{Hawking:1973uf} for the local coordinate
transformation needed to exhibit the flatness of the Misner metric \eq{dsmisner}.}
\be \label{dsmisner}
ds^2= 2 dx d\phi + x d\phi^2\,.
\ee
This corresponds to the geodesic Hamiltonian 
multiplied by a factor two compared to Eq. \eq{p2})
\be \label{Hmisner}
H(x,\phi;p_x, p_\phi) = 2 p_x  p_\phi- x  p_x^2\,.
\ee
Note that the metric and the Hamiltonian stay regular as one crosses the line $x=0$ (which is a
Killing horizon). The determinant of the metric is everywhere equal to $-1$,  and the signature is globally $- +$, 
so that the Hamiltonian is not positive definite, representing locally a difference between two squares.
In other words,  it involves an ordinary and a ghost degree of freedom.

The equations of motion read
\be
\dot x &=& 2 p_\phi - 2 x p_x\,, \nn
\dot \phi &=& 2 p_x \,, \nn
\dot p_x  &=& p_x^2 \,, \nn
\dot p_\phi &=&0\,.
\ee
Denoting  here the Hamiltonian time variable as $t$ (i.e. $\dot x = \frac{d x}{dt}$), it is simple to
solve the equations of motion. We see that $ p_\phi$ is an integral of motion. Assuming that the
initial value of $p_x$ is not zero,  
say $p_x(0) =  1/c$, the equation of motion for $p_x$
is easily integrated,  giving
\be
p_x(t)= \frac1{c-t}\,.
\ee
This leads to a blow-up for $p_x(t)$ at the finite time $t=c$, which can be positive or negative.
Correspondingly, $\phi(t)$  blows up logarithmically as $ t \to c$ 
 according to 
\be
 \phi(t) = \phi(0)- 2 \ln \left( 1-\frac{t}{c} \right)\,,
\ee
assuming $t/c < 1$.
While $p_\phi$  
stays constant,  the time-evolution of the remaining phase space variable $x(t)$
is conveniently obtained by using the constancy of the energy: $E = H(x,\phi;p_x, p_\phi)$. This yields
\be
x = \frac{2 p_x p_\phi -E}{p_x^2}\,.
\ee 
When one approaches the blow-up (where $p_x \to \infty$), $x$ tends to zero.

One can visualize the Misner manifold as a cylinder (with $\phi$ being the angular coordinate,
and $x$ the coordinate along 
 its axis). The geodesic incompleteness of the Misner manifold
means that most geodesics [apart from the ones 
 with $p_x(0)=0$] spiral [either 
for $t<0$ or $t>0$, depending on the sign of $p_x(0)$] towards
the horizon circle $x=0$ by making infinitely many turns within a {\it finite} (affine) time.
 This exemplifies how the nonlinearity of the ghost-ridden Hamiltonian \eq{Hmisner} leads to a finite-time blow-up in phase space, i.e.  is a system with a {\it malignant} ghost.

\section{Modified Korteweg-de Vries equation as a benign higher-derivative model.} \label{sec4}
 \setcounter{equation}0
        
We started our discussion with the Pais-Uhlenbeck oscillator \p{LPU}, a higher-derivative model with  
benign ghosts. We noted in the Introduction that the ghosts generically cease to be benign if a nonlinear 
interaction term  is added to the Lagrangian \p{LPU}.  The ghost models considered in Sec. \ref{sec2}
were benign essentially because the ghost degrees of freedom satisfied  linear equations of motion.
The ghost models considered in Sec. \ref{sec3} involved nonlinear evolution equations, and could
be benign (or not) depending on the global geometric properties of the considered Lorentzian manifold.
However, they were not models linked to higher-derivative Lagrangians.

A natural question at this stage  is: 
        {\it Are there nonlinear higher-derivative models with benign ghosts?}
In Ref.\cite{Toda-ja}, one of us suggested that the usual 2-dimensional $(t,x)$ 
Korteweg-de Vries (KdV) system, with the roles of
temporal and spatial variables interchanged 
might be benign because of the existence of infinitely many local
conservation laws. Indeed, let us rename  
\be 
\lb{rename}
t \to X,  \qquad x \to T\,.
\ee 
Then the local flux conservations, $0= \partial_t J^t_n+ \partial_x J^x_n \, { \equiv} \,
\partial_T J^x_n+ \partial_X J^t_n$, 
{imply} the $T$-conservation of the fluxes
\be \label{Fn}
{\cal F}_n=\int dX \left[J^x_n\right]_{T=cst} \, {\equiv} \, \int dt \left[J^x_n\right]_{x=cst}\,.
\ee
Let us recall that the 
ordinary KdV equation for $u(t,x)$, namely\footnote{As usual, we denote the
partial derivatives of $u(t,x)$ by subscripts; e.g. $u_x \equiv \partial u/\partial x$.}
 \be  \lb{KdV}
    u_{xxx} + 6 u u_x  + u_t   = 0\,,
\ee
derives from the  action $\int dt dx\,L[\psi(t,x)]$
with the 2-dimensional Lagrangian density
\be \label{LKdV}
L[\psi(t,x)] \ =\ \frac 12 \psi_{xx}^2 - \psi_x^3 - \frac 12 \psi_t \psi_x\,
\ee
if one denotes $u(t,x) \equiv \psi_x$ after having varied {over $\psi(t,x)$}.

When changing the names of spacetime variables according to Eq.\p{rename}
and denoting 
{$\psi(t,x) = \psi(X,T)  \to  \Psi(T,X)$}, the action  reads
$\int dT dX L[\Psi(T,X)]$ with a Lagrangian density
\be  \lb{LKdV-rotate}
 L[\Psi(T,X)] \ =\ \frac 12 \Psi_{TT}^2 - \Psi_T^3 - \frac 12 \Psi_T \Psi_X\,,
\ee
which contains higher-order $T$-time derivatives. The  corresponding equation 
 of motion
\be \lb{KdV-rotate}
      u_{TTT} + 6 u u_T  + u_X   = 0\,,
\ee
is of  third order in the $T$-time derivative of  $u = \Psi_T$. {As we will 
 see below,}
higher-order 
$T$-derivatives
{ in} Eq. \eq{KdV-rotate} { bring about} 
exponential instabilities  when considering the evolution in the $T$ (i.e. $x$) direction.

In order to {express} 
{our} results on the KdV {and modified KdV} equations in a  more transparent { way} 
we willl {\it not}  henceforth   use the notation $T$ for $x$ and  $X$ for $t$. We {will} simply
consider a  
non-standard Cauchy problem for the 
{equations \eq{KdV}, \eq{MKdV}}. Namely, instead of setting
the 
initial 
 value of $u(t,x)$ at\footnote{As the KdV equation is invariant under $t$ translations
(and $x$ translations), one can fix the initial-$t$ Cauchy hypersurface at $t=0$ (and the initial-$x$
Cauchy hypersurface at $x=0$).} $t=0$ { [}i.e. giving one function of $x$, $v(x)$, with the condition 
$u(0,x)=v(x)${]}, we shall now pose a ``rotated" Cauchy problem on the $x=0$ axis. However, as
Eq. \eq{KdV} features the third $x$-derivative of $u$, we must now give  {as Cauchy data {\it three independent} 
functions of $t$, $u_0(t)$,  $u_1(t)$ and $u_2(t)$, }
 with the
three conditions
\be \label{cauchy}
u(t,0)= u_0(t) \ , \ u_x(t,0)= u_1(t) \ ,  \ u_{xx}(t,0)= u_2(t).
\ee
Note that this means that we are now considering more general solutions of the KdV
equation. Indeed, any solution of the 
{ordinary} Cauchy problem {[}determined by
one function $v(x)= u(0,x)${]} will evolve {in}
$t$ (in both directions, $t>0$ and $t<0$) and will
induce on the $x=0$ axis 
some  values for the three functions $u(t,0)$,  $u_x(t,0)$ 
and $u_{xx}(t,0)${,} which are not functionally independent
 because they are all
determined by the single function $v(x)$.

The 
{ordinary} ($t$-direction) Cauchy problem for the KdV equation \eq{KdV} has been 
 shown to be {\it globally} well-posed when the single Cauchy datum $v(x)$ belongs
 to suitable functional spaces, namely $H^s$ Sobolev spaces with $s\geq1$
 (see \cite{Kenig1991} and references therein).
However, the fact that the KdV equation has good dynamic behaviour when evolved
in the $t$ direction
does not mean  that {the same is true for} its evolution in the $x$ direction.

Indeed, we 
{will} explicitly see below that the $x$ evolution gives rise {[}when considering the
linearized approximation of \eq{KdV}{]} to exponentially growing  modes (which are absent when
considering the $t$ evolution). These exponentially growing modes amplify any small-scale    
structure present in the $x$-evolution Cauchy data $u_0(t)$,  $u_1(t)$ and $u_2(t)$ 
and are not tamed by the existence of the infinitely many conserved fluxes ${\cal F}_n$ {defined in} Eq. \eq{Fn},
because their integrands $J^x_n$ are not positive definite.

Worse, Eq. \eq{KdV} admits approximate
(as well as exact) solutions which blow up along a line located at some finite distance
in the $x$ direction. 
Indeed, if one looks for power-law singularities of
$u(t,x)$ of the general form $u(t,x) \approx C(t) [x-x_0(t)]^\alpha$, it is easily seen that
they must necessarily be of the form
\be \label{blowupKdV}
u(t,x)^{\rm KdV}_{\rm blowup} \ \approx\  - 2 \left[ x-x_0(t) \right]^{-2}\,,
\ee
with an arbitrary possible blow-up line $x=x_0(t)$. A specific example of such a blow-up solution
is the following exact $t$-independent solution of the KdV equation: 
\be
      \lb{runaway}
      u(t, x) \ =\ -\frac 2{(x-c)^2} \,,
\ee
where $c$ is any constant. The function \eq{runaway} has smooth Cauchy data on the $x=0$ axis,
and blows up on the
line  $x=c$. We also performed some numerical simulations (with periodic Cauchy
data given at $x=0$) and found that many such initial data  develop a singularity 
of the form \eq{blowupKdV} during their $x$-direction evolution. This seems to confirm
the standard ``Ostrogradsky-ghost" lore that the evolution of a higher-derivative model,
such as \eq{LKdV-rotate} becomes singular during their evolution.
And one cannot expect that the quantum problem for the system \p{LKdV-rotate} would be benign.

However,
the situation {appears to be} much better for the {\it modified}\footnote{Modified
KdV equations are generally defined by replacing the nonlinear KdV term $6 u u_x$ by $n(n-1) {\kappa} u^{(n-2)} u_x$. The usual KdV equation is the case $n=3$. In this case{,} a rescaling of the variables can set the
coefficient ${\kappa}$ to $1$. Here we consider the next modified KdV equation for $n=4$. In that case{,}
a rescaling of the variables can set the coefficient ${\kappa}$ either to $1$ or $-1$; these two cases having different physical properties {--- the so-called {\it focusing} and {\it defocusing} cases}.} KdV equation,       
    \be
     \lb{MKdV} 
    u_{xxx} + 12 {\kappa} u^2 u_x  + u_t   = 0 \,.
      \ee
      This equation admits an infinite number of  integrals of motion, as the ordinary KdV equation does. The first three local conservation laws are 
       \be
       \lb{protoEnergy}
       \partial_t u = - \partial_x(u_{xx} + 4{\kappa} u^3)\,,
       \ee
       \be
       \lb{Energy}
       \partial_t u^2 = -2\partial_x\left[3{\kappa} u^4 + u u_{xx} - \frac 12 u_x^2 \right]\,,
       \ee
       
       \be
       \lb{superEnergy}
       \partial_t\left({\kappa} u^4 - \frac 12 u_x^2 \right) \ =\ \partial_x\left[{\kappa} u_x(12 u^2 u_x + u_{xxx})
       - \frac 12 u_{xx}^2 - 4 {\kappa} u^3 u_{xx} - 8 {\kappa}^2 u^6  \right]\,.
        \ee 
      {We will argue in the following that, in constrast to Eq. \p{KdV}, the equation \p{MKdV} does not involve a blow-up}.

Let us first briefly discuss the linearized KdV equation,
\be \lb{lKdV} 
 u_{xxx} + u_t   = 0 \,,
 \ee
{which describes the fluctuations} around the
trivial solution $u(t,x)=0$
of the usual 
 KdV equation \eq{KdV}, as well as 
 of all its modified versions with the nonlinear term $n(n-1) \kappa u^{(n-2)} u_x$, $n \geq 3$.
 
 Let us
{emphasize} the drastic difference between the $t$-evolution Cauchy problem,
 and the $x$-evolution Cauchy problem of the linearized equation \eq{lKdV}. 
 Equation \eq{lKdV} can be solved by decomposing the solution $u(t,x)$ 
 in  plane waves $ e^{ {\rm i} (\omega t + k x)}$.  
This yields
 the dispersion law
 \be \lb{disp}
 \omega = k^3\,.
 \ee
 If one poses the usual Cauchy problem 
with some Fourier-transformable initial data
 \be
 u(0,x) = v(x) \equiv \int \frac{d k}{2 \pi} \, v(k) e^{ {\rm i}  k x}\,,
 \ee
 the $t$-evolution of the initial data $v(x)$ yields the solution (valid for both signs of $t$)
 \be
 u(t,x) = \int \frac{d k}{2 \pi} \, v(k) e^{ {\rm i}(k^3t+  k x)}\,.
 \ee 
 The important point here is that $u(t,x)$ is obtained from $v(k)$ by a purely oscillatory complex
 kernel  $e^{ {\rm i}(k^3t +  k x)}$ of unit modulus. It has been shown that this oscillatory
 kernel has {\it smoothing} properties (see, e.g., \cite{Kenig1991}). This allows one to take
 the initial data in low-$s$ Sobolev spaces $H^s$ (describing pretty rough initial data) \cite{Kenig1991}.
 
{On the other hand}, if one considers the $x$-evolution Cauchy problem, one starts from three independent
functions of $t$ along the $x=0$ axis: $u(t,0)= u_0(t)$,  $u_x(t,0)= u_1(t)$ and $u_{xx}(t,0)= u_2(t)$, {see}
Eq. \eq{cauchy}.
Assuming that the three Cauchy data $u_a(t)$, $a=0,1,2$, are Fourier-transformable, we can represent
them as
 \be
 u_a(t)  \equiv \int \frac{d \omega}{2 \pi} \, u_a(\omega) e^{ {\rm i}  \omega t}\,.
 \ee
 The three Cauchy data determine a unique solution which, when decomposed
 in plane waves, satisfies the same dispersion law \eq{disp} as before. However,
 the dispersion law \eq{disp} must now be solved for $k$ in terms of  $\omega$.
 As it is a cubic equation in $k${,} it has {\it three different roots}, namely 
 \be
 k_a(\omega)= \omega^{\frac13} j^a \,,
  \ee
where $\omega^{\frac13}$ denotes the unique real cubic root of $\omega$ 
and where $j^a =  1, j, j^2$  (with $ j \equiv e^{ \frac{2 \pi \rm i}{3}}=-\frac12 + {\rm i} \frac{\sqrt{3}}{2}$) 
 {are}
the three complex roots of unity. This yields a solution for $u(t,x)$ of the form
\be
u(t,x)=\sum_{a=0,1,2} \int \frac{d \omega}{2 \pi} \, v_a(\omega) e^{ {\rm i} ( \omega t+ k_a x)}\,,
\ee
where the three coefficients $v_a(\omega)$ are (uniquely) determined by the three initial conditions
at $x=0$, namely by the following system of three linear equations\footnote{The determinant of this system is $- 3 \sqrt{3} \omega$.}
\be
u_0(\omega) &=& \sum_{a=0,1,2} v_a(\omega)\,, \nn
u_1(\omega) &=& {\rm i} \sum_{a=0,1,2}k_a(\omega) v_a(\omega)\,, \nn
u_2(\omega) &=& - \sum_{a=0,1,2} k_a^2(\omega) v_a(\omega)\,.
\ee
The point of this exercise was to exhibit the fact that, when considering the $x$ evolution with arbitrary
Cauchy data $u_0(t)$,  $ u_1(t)$, $u_2(t)$, the solution involves {\it exponentially growing} modes
in the $x$ direction, linked to the fact that ${\rm i} k_1= {\rm i} \omega^{\frac13} j$  and 
${\rm i} k_2= {\rm i} \omega^{\frac13} j^2$ have {\it real} parts $\pm \frac{\sqrt{3}}{2} \omega^{\frac13}$
(this holds for both signs of $x$). 
{This instability is associated with the presence of higher time derivatives.}\footnote{{We hasten to comment, however, that higher derivatives do not {\it necessarily} entail such an instability. For example, it does not show up in the equations of motion for the Pais-Uhlenbeck system \p{LPU}.} }

 From the mathematical point of view, this means that high-frequency (HF) wiggles, in the sense
of high $\omega$,  in the initial Cauchy data (at $x=0$) will be  amplified, on both sides of the $x$ axis,
by the exponentially growing  factor $e^{\frac{\sqrt{3}}{2} |\omega|^{\frac13} |x|}$. 
This indicates that the Cauchy problem will be well-posed
only if one takes initial data whose Fourier transforms $v_a(\omega)$ decrease sufficiently
fast as $|\omega| \to +\infty$. As a minimum condition for a local existence theorem, one 
 should require the Fourier transforms $v_a(\omega)$ to decrease like 
$e^{- c |\omega|^{\frac13} }$ for some positive constant $c$. 
 This essentially defines the $s=3$ Gevrey class of 
functions on ${\mathbb R}$ (see, e.g.,\cite{Rodino}). Note that such a regularity condition is stronger than infinite differentiability,
but weaker than analyticity (which corresponds to a decrease of the Fourier transforms
of the type $e^{- c |\omega| }$, i.e. to the $s=1$ Gevrey class).

We therefore expect that it will be mathematically possible to prove (at least for sufficiently weak Cauchy data)
a {\it local} existence theorem for solutions of the $x$-evolution of the modified KdV equation \eq{MKdV} 
in suitable Gevrey classes  (say, with $1<s \leq 3$).\footnote{ Nader Masmoudi confirmed (private communication to TD) that it was likely 
that taking Cauchy data in the $s=3$ Gevrey class would suffice for local existence of the solution of
the $x$ evolution. This would still leave open the issue of finding adequate function spaces for global
existence.}

We wish to further argue that the special nonlinearity of Eq. \eq{MKdV}, when taking ${ \kappa} >0$,
is likely to allow sufficiently smooth {$x=0$}  
Cauchy data  to define {\it global solutions}, extending to
arbitrary large values of $|x|$. 
Our main argument for believing that the local $x$-evolution of Eq. \eq{MKdV} can be extended to large values of $|x|$
is the absence of blow-up solutions.  Indeed, let us look for power-law singularities of
$u(t,x)$ of the general form 
\be \label{blowup}
u(t,x) \approx \beta(t) {[} x-x_0(t) {]} ^\alpha \,,
\ee
with $\alpha <0$. Inserting such an asymptotic behavior in Eq. \eq{MKdV}, it is easily seen that
the $t$-derivative term $u_t$ is necessarily subdominant with respect to the $x$-derivative terms. 
Therefore, blow-up solutions must, in  lowest approximation, solve the truncated equation
\be
     \lb{MKdVs1} 
 0=   u_{xxx} + 12 {\kappa} u^2 u_x  \ =\  \frac{\partial}{\partial x} (u_{xx} + 4 {\kappa} u^3)   \,.
 \ee
 This equation is a third-order dynamical equation for the $x$ evolution which admits the ``constant of motion"
 \be
   u_{xx} + 4 {\kappa} u^3 \ = \ C \,.
 \ee
 In turn, the latter equation can be rewritten as $ u_{xx}= - \frac{d V(u)}{d u}$,
 i.e. as the Newtonian  equation of motion in $x${, playing the role of time,} for 
 a particle with position $u$ in the following potential: 
         \be
         \lb{poten}
       V(u) = {\kappa}  u^4 - C u\,.
         \ee
When ${\kappa} >0$ {(the focusing case)}, this potential grows for large values of $|u|$, and therefore prevents the existence of blow-up 
solutions.\footnote{By contrast, for the ordinary KdV, one obtains a non-confining cubic potential, and 
the corresponding equations of motion admit the singular  run-away solutions \p{runaway}. 
General modified KdV equations with the nonlinearity $n(n-1) {\kappa} u^{(n-2)} u_x$ give rise to a potential $V(u)= {\kappa} u^n - C u$, which is confining if $n$ is even and 
${\kappa} >0$. }
Technically, if we look for blow-up solutions of the type 
\eq{blowup}, one finds that the blow-up exponent $\alpha$
must be equal to $\alpha=-1$ and that the coefficient $\beta$ must satisfy the cubic equation
$ \beta (2 {\kappa} \beta^2 +1)=0$. When $ {\kappa} <0$ {(the defocusing case)}, there exist  
possible non-trivial (real) blow-up solutions
with $\beta= \pm (- 2 {\kappa})^{- \frac12}$. [These solutions correspond to the ``fall" of the $u$ particle down the
unstable quartic potential ${-|\kappa|}  u^4 - C u$.
] By contrast, when ${\kappa} >0$, the only real solution of the
cubic equation of $\beta$ is $\beta=0$. In other words, there exists no real blow-up solutions of the 
type
indicated in Eq. \eq{blowup}.\footnote{We are aware that the ansatz \eq{blowup} might be too restrictive. {It was shown in Ref. \cite{Merle}} that, in the formally confining
case $n=6$ with ${\kappa} >0$, a blow-up can occur in the $t$-evolution for data
close to the corresponding soliton.}

A different (though related) analytic argument indicating the absence of real blow-up solutions
comes from the analysis of the scaling properties of the modified KdV equation. It is easily seen 
that Eq. \eq{MKdV} is invariant under the rescalings $u= \lambda_u \bar u$, $x= \lambda_x \bar x$,
 $t= \lambda_t \bar t$ 
if
\be
\lambda_t = \lambda_x^3 \ ; \ \lambda_u = \lambda_x^{-1}\,.
\ee
The quantities $ x u$ and $\frac{ x}{t^{1/3}}$ are invariant under these rescalings. Using also
the space and time translational invariance of the modified KdV equation, we can 
look for scaling solutions of the type
\be \lb{scale-Ans}
u(t,x)= \frac{  \alpha}{[3(t-t_0)]^{1/3}} w(z)\,,
\ee
where
\be
z \equiv \frac {x-x_0}{[3(t-t_0)]^{1/3}} \,.
\ee
Inserting the ansatz \eq{scale-Ans} in Eq. \eq{MKdV} 
 and choosing the normalization
constant $\alpha=\frac1{\sqrt{2}}$, it is easily checked that the function $w(z)$
must satisfy the equation
\be
0=w'''+ (6 {\kappa} w^2 -z) w'- w=\frac{d}{dz} \left[  w'' + 2 {\kappa} w^3 - z w  \right]\,.
\ee
Denoting as $ C$ the constant value of the bracket in the last right-hand side, we conclude that
$w(z)$ satisfies a second-order equation of the type
\be \label{P2}
 w'' = - 2 {\kappa} w^3 + z w  + C \,.
\ee
This is a Painlev\'e II equation \cite{Pain}. In general, Painlev\'e equations  have (moving) pole singularities. And indeed,
a local analysis of Eq. \eq{P2} (keeping the leading-order terms\footnote{Note in passing that
this leading-order equation describe the dynamics of a particle in the potential $V(w)=\frac12 {\kappa} w^4$}
 $ w'' \approx - 2 {\kappa} w^3$)
shows that \eq{P2} admits simple poles, $w(z)= R/(z-z_0)$, as local singularities if the residue $R$
satisfies the equation ${\kappa} R^2=-1$. When ${\kappa} <0$, the residue $R$ will be real, so that real solutions
of Painlev\'e II can have (and generally do have) poles on the real $z$ axis. The existence of a
real simple pole at $z=z_0$ would then correspond to a singular (blow-up) behavior of $u(t,x)$ of the form 
$u(t,x) \propto \left[x-x_0 - z_0  [3 (t-t_0)]^{1/3} \right]^{-1}$. However, when ${\kappa} >0$,
real solutions cannot have real poles. This excludes the existence of (real) singular scaling solutions.

Another way to understand why the negative-${\kappa}$ modified KdV equation has blow-up solutions
is to use its relation with the usual KdV equation \eq{KdV} (which admits real blow-up solutions,
as we emphasized above). Indeed, it is well-known that the Miura transformation,
         \be
         \lb{Miura}
          u \ =\ -(2v^2  + \sqrt{2} v_x)\,,
           \ee
  transforms the ordinary  KdV equation \eq{KdV} for $u(t,x)$ into  the equation
     \be
     v_t  - 12v^2 v_x  +  v_{xxx}\  = \ 0 \,,
      \ee
which coincides with the modified KdV equation \eq{MKdV} for $v(t,x)$, with ${\kappa} =-1$.

\begin{figure} [ht!]
      \begin{center}
    \includegraphics[width=.9\textwidth]{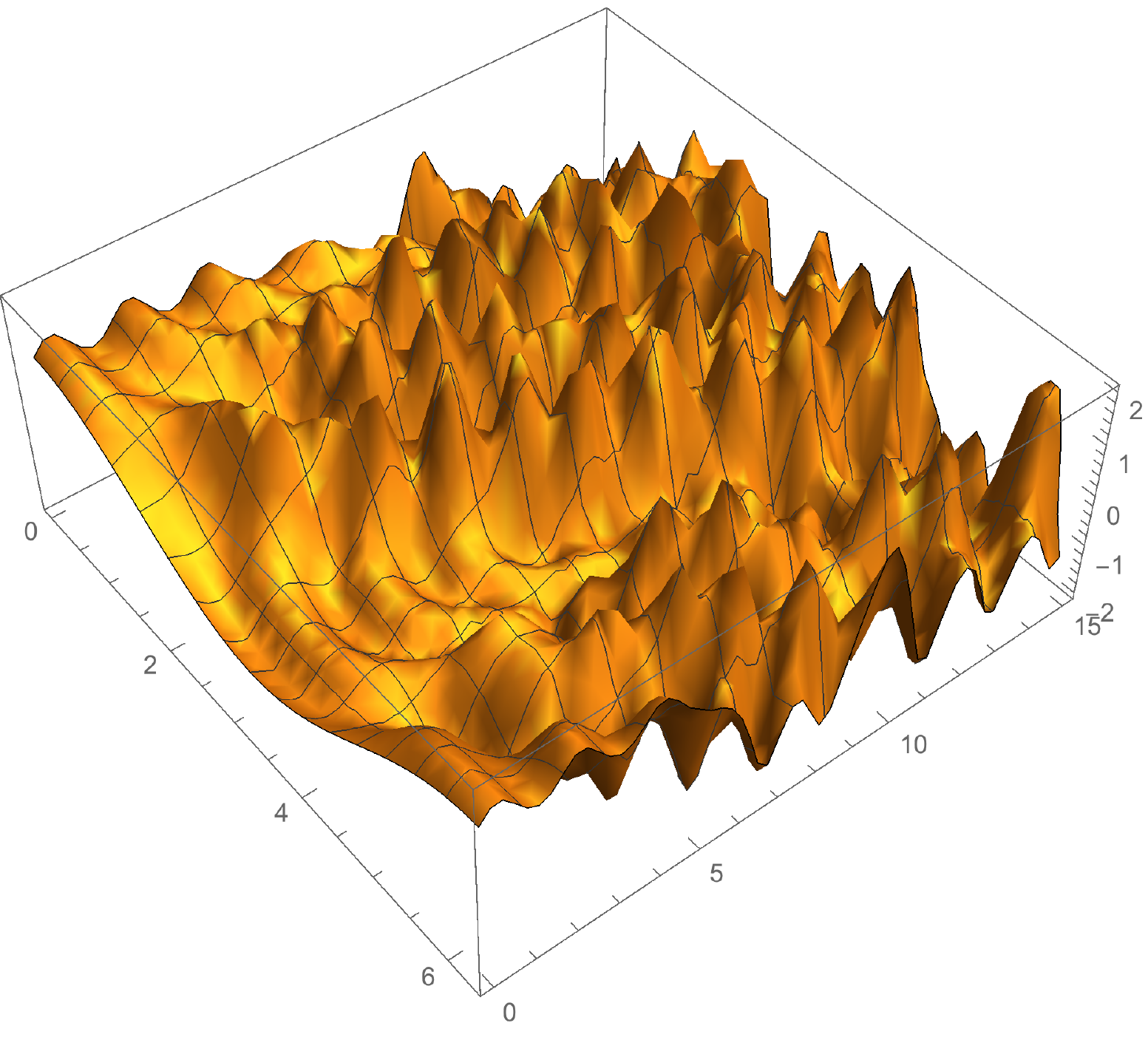}                  
     \end{center}
    \caption{Solution $u(t, x)$ of the $\kappa=+1$ modified KdV equation \eq{MKdV} corresponding to the $t$ periodicity
    \eq{period} and the $x=0$ Cauchy data \eq{cauchynum}.}        
 \label{solMKdV}
    \end{figure}

{To confirm} our conjecture that the $x$-evolution of sufficiently smooth  Cauchy data  
\eq{cauchy}
stays bounded when evolved with the modified KdV equation with ${\kappa} >0$, we have performed some numerical
simulations (done with {\it Mathematica}) for the case ${\kappa}=+1$. To simplify the numerical analysis we imposed {[}as is allowed by Eq. \eq{MKdV}{]}
periodicity along the $t$ direction 
{Using} scaling invariance, we can assume 
$2 \pi$ periodicity:
 
\be
        \lb{period}
        u(t + 2\pi, x) \ =\ u(t,x) \,.
\ee
In order to study the effect of the nonlinear term $ 12u^2 u_x${,}  we used $t$-periodic Cauchy data
for which the nonlinear term is initially (i.e. at $x=0$)  larger than the linear term $u_t$. In particular, we used
the Cauchy data
\be \lb{cauchynum}
  u(t,0) \ =\ \cos t, \ u_x(t, 0) = \cos t, \ u_{xx}(t,0) \ =\ 0\,.
\ee
We first checked that the use of such Cauchy data for the modified KdV equation {with  ${\kappa}=-1$ } was leading quite
fast  (namely at $x=0.887717$) to a singularity. By contrast, our numerical simulations of the $x$ evolution
of the ${\kappa}=+1$ modified KdV equation showed that $u(t,x)$ stayed bounded for all the values of $x$
that we explored. This is illustrated in Fig. \ref{solMKdV} which displays  the solution $u(t,x)$ generated
by the Cauchy data \eq{cauchynum} in the domain $0 \leq t \leq 2\pi$, $0 \leq x \leq 15$.

Fig.  \ref{solMKdV} illustrates the benign nature of the ghostful $x$-dynamics of the modified KdV equation
in the positive{-$\kappa$} case. The presence (when neglecting the term $u_t$) of an approximate $x$ dynamics
governed by the confining potential \eq{poten} reflects itself in the oscillations in the $x$ evolution
of $u(t,x)$, i.e. in the ``stormy-sea" aspect of  $u(t,x)$ in the $x >5$ part of Fig.  \ref{solMKdV}.
Note that we have taken here analytic ($C^{\omega}$) data, which generate (at least locally) an analytic solution.
We leave to future work to clarify how less regular Cauchy data  (e.g. taken in Gevrey classes, or suitable Sobolev-type
spaces) would evolve under the $x$ evolution. 

We expect generic Cauchy data \eq{cauchy} for the $x$ evolution to 
evolve into stormy-sea solutions similar to the one illustrated in Fig.  \ref{solMKdV}. However,
there will also exist special (measure-zero) Cauchy data that will evolve into much tamer solutions.
Indeed, if we start with a smooth function of $x$, say $v(x)$, and use it as unique Cauchy datum at $t=0$ {[}namely, $u(t=0,x)=v(x)${]}, its $t$-evolution will define a smooth solution 
$u^{v \,\rm sol}(t,x)$. The restriction
to the $x=0$ axis of  $u^{v \,\rm sol}$,  $u^{v \,\rm sol}_x$, and  $u^{v \,\rm sol}_{xx}$ will then define
Cauchy data for the $x$-evolution {that generate} the smooth solution $u^{v \,\rm sol}(t,x)$.
We have numerically checked this fact by using solitonic solutions of the modified KdV equation.

Let us recall that, like the ordinary KdV equation, the modified equation \p{MKdV} admits solitonic  solutions. 
Indeed, one can look for travelling-wave solutions $u(t,x)= u(x+ c t)$, moving with some celerity $c$, by inserting the
ansatz  $u(t,x)= u( \bar x)$, with $\bar x \equiv x+ c t$, in Eq. \eq{MKdV}. It is easily seen that this yields
the equation
\be
\frac{\partial}{\partial \bar x} \left[ c u+ 4 {\kappa} u^3 + u_{\bar x \bar x}  \right]=0\,.
\ee
Denoting as $C_0$ the constant quantity within the bracket, we then get the following second-order equation
for the function $u(\bar x)$:
\be
 u_{\bar x \bar x}= - \frac{d}{d u} {\cal V}(u)\,,
\ee
with a potential function ${\cal V}(u)$ now given by
\be
{\cal V}(u)= {\kappa}  u^4 + \frac12 c u^2 - C_0  u\,.
\ee
We are again reduced to the dynamics of a particle moving in the confining
quartic potential
${\cal V}(u)$, considered for some fixed ${\kappa} >0$ (say,  ${\kappa}=+1$) . 
The general solutions of this problem then depend 
on three parameters: the celerity $c$, the constant $C_0$, and the constant energy of the $\bar x$ dynamics:
\be
E = \frac12 u_{\bar x}^2 + {\cal V}(u)\,.
\ee
The usually considered solitonic solutions (such that $u(\bar x)$ tend to zero when $\bar x \to \pm \infty$)
are obtained by taking $c <0$, $C_0=0$ and $E=0$. This gives a symmetric double-well potential: $  {\kappa} \, u^4 - 
\frac12 |c| \, u^2$. The zero-energy solution then  describes a $u$-motion which starts, at ``time"$\bar x = - \infty$, 
at $u=0$ with zero ``velocity" $u_{\bar x}$,
 {glides down (say) to the right, reflects on the right wall of the double well and then turns back}
to end up again at  $u=0$ when  $\bar x = + \infty$. The explicit form of the corresponding solution is 
\be 
{u({\bar x}) \ =\ \sqrt{\frac {|c|} {2{\kappa}}} \, \frac 1 {\cosh (\sqrt{|c|} \,\bar x) } \,.}
 \ee

Here, we are interested in constructing {\it periodic} travelling-wave solutions 
satisfying $u(\bar x + \bar T)= u(\bar x)$
for some $\bar T$. Such solutions can be easily constructed by considering bound mechanical motions {in the potential ${\cal V}(u)$  having a 
{\it non-zero energy}}. We have already given such oscillatory solutions in a 
(symmetric) quartic potential (i.e. for $C_0=0$) in Eq. \eq{cn} above.
The corresponding periodic travelling-wave soliton  is then of the form
\be
 u(\bar x) \ =\ x_0 \, {\rm cn} [\Omega(\bar x- {\bar x}_0), k
]\,,
 \ee
as simply obtained by using in Eqs. \eq{V-Robert}, \eq{cn}, \eq{cnparams} the replacements
\be
 x \ \to \  u, \; t \to \bar x, \; \omega^2 \ \to \  c ,\;   \lambda \ \to \ 4 {\kappa}, \; N \ \to E \,.
 \ee

  We have used the $t$-periodic Cauchy data defined
 by restricting $u(x+ c\,t)$, and its first two $x$ derivatives, to $x=0$ to check the  accuracy of our
numerical simulations. The numerical solution  generated from these Cauchy data agreed well with the
analytical solution for values of $x$ of order $\bar T$. For larger values they exhibited some numerical noise.
The specific form of the numerical noise depended on the numerical scheme used. When using the same scheme
as the one used to produce Fig.  \ref{solMKdV}, the noise stayed at low frequencies. This gives us confidence
that Fig.  \ref{solMKdV} yields a reasonably accurate picture of the benign nature
of the $x$-evolution of smooth Cauchy data.

       \section{Discrete nonlinear  systems with benign ghosts.}
        \setcounter{equation}0
       
   In this final section, we define some higher-derivative dynamical models 
having only a finite number of degrees of freedom   
and exhibiting  a benign behavior in their evolution. These models are constructed
   by discretizing the modified KdV equation in the $t$ direction, keeping continuous the $x$ evolution.
   [We recall that $x$ is our timelike evolution variable.] This will replace the partial differential equation
   \eq{MKdV} by a system  of coupled ODEs with respect to $x$.

Similarly to the derivation of the ordinary KdV equation  from the Lagrangian \p{LKdV},  the modified KdV equation \p{MKdV} follows from the two-dimensional action
   $S = \int dt dx L[\psi(t,x)]$, with the field-theory Lagrangian density
      \be
      \lb{LMKdV}
      L[\psi(t,x)] \ =\ \frac 12 \psi_{xx}^2 - {\kappa} \psi_x^4 - \frac 12  \psi_x \psi_t\,.
       \ee
 After varying the action with respect to $\psi(t,x)$, one gets Eq.  \p{MKdV} by substituting $ \psi_x(t,x) \to u(t,x)$.
       
 We can then define a discretized version of the  modified KdV dynamics by
 assuming that the variable $t$ takes only the discrete values $t=h, 2 h, \cdots, N h$, 
for some integer $N \geq 2$,
 and by replacing the continuous time derivative $\psi_t$ by a discrete (symmetric) time derivative 
 $\frac{\psi(t+h,x)-\psi(t-h,x) }{2 h}$. This yields an action  of the form $S_N= \int dx \, L_N$,
 where the Lagrangian $L_N$ is given by a sum of $N$ terms: 
 \be \label{LN}
L_N= \sum_{t=h}^{t=N h} \left[  \frac 12 [\psi_{xx}(t,x)]^2 - {\kappa} [\psi_x(t,x)]^4 - \frac 12  \psi_x(t,x) \frac{\psi(t+h,x)-\psi(t-h,x) }{2 h}  \right]\,.
 \ee
 To define the model we further need to specify  boundary conditions. 
Namely, we need to define $\psi(0 h,x)$ and $\psi[(N+1)h,x]$, which enter the discrete $t$ derivative
 for $t=h$ and $t=N h$, respectively. This can be done in two different ways: {\it (i)} we can use Dirichlet-type boundary conditions,
 namely  $\psi(0 h,x)=0$ and $\psi[(N+1)h,x]=0$ or {\it (ii)} periodic boundary conditions,  $\psi(0 h,x)=\psi(N h,x)$,
 and  $\psi[(N+1) h,x] = \psi(h,x)$. The periodicity condition can only be 
{imposed} when $N \geq 3$. 
 
We expect that taking larger and larger values of $N$ would allow one to simulate better and better  the 
continuous theory (though the presence of chaos might make such a convergence non uniform in $x$).

 The simplest discretized model is obtained by {choosing $N=2$ and}  taking Dirichlet-type boundary conditions.
 This model has only two dynamical variables: 
       $\psi(x) \equiv \psi(h,x) $ and $\chi(x) \equiv  \psi(2 h,x)$. We recall that  $x$ is playing the role of time.
  The Lagrangian $L_2 \equiv L_{N=2}$ reads
       \be 
       \lb{Lpsichi}
       L_2 \ =\  \frac 12 \psi_{xx}^2 + \frac 12 \chi_{xx}^2 - {\kappa}\psi_x^4 - {\kappa} \chi_x^4 -\frac1{4h}\psi_x \chi+ \frac1{4h} \chi_x \psi  \,.
        \ee
   
{Adding} a total $x$-derivative, the last {two terms} can be  traded for 
$+ \frac1{2h} \chi_x \psi $.          
  Using suitable rescalings, we can set ${\kappa}=1$ and $h=\frac 12$.  For simplicity, we will use these values
  in the following.

Defining the two new dynamical variables      $a(x) \equiv \psi_x, b(x) \equiv  \chi_x$, the equations
of motion derived from the Lagrangian     \eq{Lpsichi} read
        \be
        \lb{eqmot-ab}
        a_{xxx} + 12 a^2 a_x + b  &=& 0 \,,\nn
        b_{xxx} + 12 b^2 b_x - a &=& 0 \,.
         \ee
This is a system of two coupled  higher-order evolution equations. The important point is that, while we had
technical difficulties in numerically simulating
in a stable manner 
the $x$-evolution of the (periodic) modified
KdV equation \eq{MKdV}, we could easily perform long-term integrations of the coupled system \eq{eqmot-ab}
up to $x = 10 000$ {and more} 
in a numerically stable manner. And though the conserved energy of the system, namely
\be
         \lb{Energy-ab}
         E \ &=&\  \frac 12 (\psi_{xx}^2 + \chi_{xx}^2 ) - 3(\psi_x^4 + \chi_x^4) - \psi_x \psi_{xxx} - \chi_x \chi_{xxx} \nn
        \ &=& \  \frac 12 (a_x^2 + b_x^2) - 3(a^4 + b^4) - a a_{xx} - b b_{xx} 
\ee
 [which is a discrete version of the conserved current on the right-hand side of Eq. \p{Energy}{]}, is not positive-definite
and can take arbitrarily positive or  negative values, our numerical simulations indicate that the classical motions
have a benign behavior, without any blow{-}up. In other words, the simple discrete model $L_2$, Eq. \eq{Lpsichi},
provides a nontrivial example of an {\it interacting}  higher-derivative system with benign ghosts.

          One can also write the corresponding Ostrogradsky Hamiltonian. It reads 
           \be \label{Ham-Ostr}
           H \ =\ \frac 12 (P_\psi^2 + P_\chi^2) + p_\psi \psi_x + p_\chi \chi_x + \psi_x^4 + \chi_x^4 - \psi \chi_x\,,
            \ee  
            where $\psi, \psi_x, \chi, \chi_x$ should all be treated as independent variables 
with
             corresponding canonical momenta $p_\psi, P_\psi, p_\chi$ and $P_\chi$,  respectively.
         The Hamilton equations of motion following from \p{Ham-Ostr} coincide with \p{eqmot-ab}.

  For completeness, let us also write the periodic, discrete dynamics defined by $L_N$, Eq. \eq{LN},  
{with} $N \geq3$.
{Setting} as before ${\kappa} =1$ and $h=\frac12$ 
 and defining the $N$ $x$-evolving discrete variables    
  $a^k(x) \equiv \psi(k h,x)_x$, with $k=1, \cdots, N$, and the periodicity conditions,
    $a^0 \equiv a^N$ and $a^{N+1} \equiv a^1$, the equations of motion following from Eq. \eq{LN} read
       \be 
       a^k_{xxx} + 12 (a^k)^2 a^k_x \ =\ a^{k+1} - a^{k-1}\,.
        \ee
        The conserved energy is
        \be \label{EN}
        E \ =\ \sum_{k=1}^N \left[ \frac 12 (a^k_x)^2 - 3(a^k)^4  - a^k a^k_{xx}  \right]\,.
         \ee  
     The periodic systems with $N \ge 3$ enjoy also a second integral of motion (linked to the periodicity in $t_k = k h$):   
   \be \label{QN}
 Q \ =\ \sum_{k=1}^N \left[ a^k_{xx} + 4 (a^k)^3  \right]\,.
  \ee
In the continuous theory{,} this follows from integrating over the periodic variable $t$ the current in the right-hand side
 of Eq.\p{protoEnergy}. By contrast, the currents  in the higher conservation laws of  the modified KdV equation,
 starting with Eq. \eq{superEnergy}, do not translate into integrals of motion of the discrete systems.
  
  The $N$-th periodic discrete system has only two integrals of motion, Eqs. \eq{EN}, \eq{QN}, and $2N$ {\it pairs} of phase space variables. [For $N=2$ one had only one integral of motion for {four pairs of} phase space variables.] The systems \p{LN} are thus not integrable and their trajectories are expected to exhibit a chaotic behavior. Our numerical
  simulations did confirm this expectation.

{The} numerical confirmations  of the 
 benign nature of the discrete systems up to {rather} large {values of} $x$ {can be considered} as a further argument in favour of the conjecture that
  the continous modified KdV system \p{MKdV} is also benign.

\section{Conclusions}
 \setcounter{equation}0

  We presented several nontrivial examples of higher-derivative systems including ghosts, but where the ghosts are of 
  benign nature, i.e. {they} do not lead to {a} blow-up in the classical case (and hence will not give rise to unitarity violation 
  in the quantum case). The most interesting example is the two-dimensional modified KdV system \p{LMKdV}
   (with ${\kappa} >0$),
  viewed as a higher-derivative evolution in the $x$ variable. We presented several  arguments strongly indicating
   that this system does not involve any blow-up during its $x$ evolution.  Mathematically proving that
   this is the case (for smooth enough data) is a challenge for the future. 
          
      In the last section, we presented mechanical systems, with a finite number of degrees of freedom,
     {which} are $t$-discretized avatars of the modified KdV system \p{LMKdV}. These systems
      are of interest  on  their own, notably because they provide a set of nontrivial interacting higher derivative systems with benign ghosts. Such systems were not known before. 
          
          \section*{Acknowledgements}
          We are indebted to  Piotr Chrusciel, Alberto De  Sole,  Victor Kac, Nader Masmoudi, Frank Merle, and
       Laure Saint-Raymond for informative discussions.


\end{document}